\documentclass{article}
\usepackage{graphicx} 

\usepackage{style}
\usepackage{authblk}

\title{Estimating Residential Displacement in the Central Puget Sound Region using Household Survey Data}
\author[1]{Ameer Dharamshi}
\author[2]{Mary Richards}
\author[3]{Suzanne Childress}
\author[2]{Brian Lee}
\author[4]{Daniel Casey}
\affil[1]{Department of Biostatistics, University of Washington}
\affil[2]{Puget Sound Regional Council}
\affil[3]{Metropolitan Transportation Commission}
\affil[4]{Public Health - Seattle \& King County}

\begin{document}

\maketitle

\begin{abstract} 
Housing instability is a persistent challenge faced by households in cities across the United States. In worst-case scenarios, households are displaced from their residences and forced to start anew. In an effort to mitigate the harms of residential displacement, local policymakers have an interest in monitoring residential displacement within their communities. In this work, we propose a new strategy to estimate sub-county residential displacement within the Central Puget Sound Region using data from three household survey programs. We first estimate residential displacement between 2016-2023 from a local household travel survey using a Bayesian spatiotemporal model, and poststratify with data from the American Community Survey. We then benchmark these estimates to the American Housing Survey to ensure consistency across sources. The results reveal east-west and north-south differences in residential displacement rates within the region as well as a temporary moderation of displacement in the 2020-2021 cohort of movers. Our estimates are publicly available for interested stakeholders to further study trends in residential displacement in the Central Puget Sound Region, and our methodology is transportable to other jurisdictions with similar data contexts. 
\end{abstract}

\section{Introduction}

Many members of society live in a persistent climate of housing insecurity \citep{deluca2022housing}. They are priced out of ownership, struggle to pay rent, face rising eviction rates, and/or may be displaced from their homes and forced to start anew, often in materially worse circumstances \citep{desmond2015bforced, aurand2023shortage}. 

Residential displacement has attracted substantial attention from local policymakers and scholars as they revisit housing policies. Historically, residential displacement has been studied primarily in the context of gentrification, the process by which a low-income and neglected neighbourhood is drastically changed through an inflow of high-income and/or high educational attainment individuals \citep{chapple2021gentrification}. Scholars have focused on the questions of whether gentrification leads to the displacement of pre-gentrification residents, and if so, who is displaced and to what degree; see for instance \cite{zukin2016gentrification},  \cite{easton2020measuring}, \cite{hwang2020unequal}, \cite{finio2022measurement}, and \cite{freeman2024they}. These studies typically focus on a subset of neighbourhoods identified as gentrifying, and examine changes in migration flows as well as demographic, social, and economic characteristics of the neighbourhoods and movers under study.

\cite{zuk2018gentrification} argued that residential displacement, while an important part of the gentrification story, is ``analytically distinct"; that is, many households are at risk of displacement for reasons other than gentrification and in areas other than gentrifying neighbourhoods. Understanding the dynamics of residential displacement beyond its role as a downstream effect of gentrification is needed to inform policies to support households both at risk of displacement and those that have already been displaced.

Despite the demand, metrics of residential displacement are not readily available. The primary barrier has been measurement: how exactly should residential displacement be defined and what suitable data sources are available to quantify it? The standard definition derives from \cite{grier1978urban}; they formally define residential displacement as having occurred when a ``household is forced to move from its residence by conditions which affect the dwelling or immediate surroundings, and which: 1. are beyond the household's reasonable ability to control or prevent; 2. occur despite the household's having met all previously imposed conditions of occupancy; and 3. make continued occupancy by that household impossible, hazardous or unaffordable." This definition is broad, encompassing those displaced by gentrification as well as those displaced by other factors such as ``renevictions", untenable rent increases, eminent domain, and so on. 

Quantifying residential displacement according to this definition has remained challenging, largely due to a lack of appropriate data as movers are not asked about their reasons for moving as a matter of routine \citep{chapple2017developing, easton2020measuring}. For instance, the American Community Survey (ACS), the premier data source on the economic, social, and demographic state of American life contains information on housing tenure (i.e., it identifies movers), but it does not contain direct information on residential displacement. In practice, displacement is primarily measured via dedicated submodules within housing-related household surveys. The American Housing Survey (AHS) is the primary such product: it asks recent movers (i.e., those that moved in the previous two years) why they moved, which can be recoded into an indicator for displacement, and then reports the results at the level of large metropolitan areas (specifically, for the most populous core-based statistical areas). 

But what about measurements of local displacement? City planners and policymakers are often interested in understanding how residents are moving \emph{within their jurisdictions}. The lack of geographic granularity in the AHS limits its utility for this task. To address this challenge, two primary approaches have emerged. In some cities, dedicated local housing surveys contain similar specialized displacement questions that can be used in place of the AHS: for example, \cite{desmond2015forced}, and \cite{freeman2004gentrification}, \cite{wyly2010displacing}, and \cite{carlson2020measuring} use such data to study the drivers and effects of residential displacement in Milwaukee and New York City, respectively. This type of data is not available broadly as large scale housing surveys are challenging projects for cities; they are expensive to run and require significant expertise on survey design and implementation. In the absence of direct data, urban planning researchers have developed proxy measures of residential displacement using composite indices of risk factors. The Urban Displacement Project, for example, produces tract-level maps of estimated displacement risk indices and typologies for a number of major urban centres, both in the United States and around the globe \citep{chapple2021urban}. This work offers insights into the risk factor profiles of cities, but interpretability is challenging as it is difficult to disentangle the precise drivers of each index or to define how these indirect indices relate to displacement itself; especially when the relationship between risk factors and displacement is complex and spatially heterogeneous.

In this paper, we study residential displacement in the Central Puget Sound Region, defined here as the Census designated Seattle-Tacoma-Bellevue metropolitan statistical area, which contains the combined area of King, Pierce, and Snohomish counties in Washington state. There are several competing factors influencing housing in the region including: 1. Seattle, the largest city in the Central Puget Sound Region, and several nearby suburbs have experienced a massive tech boom which has triggered a realignment of the housing market through the influx of a large number of high-income tech workers; 2. the degree of sprawl is limited by the natural bodies of water in the region as well as the Washington State Growth Management Act, which restricts development outside designated Urban Growth Areas; 3. the legacy of historical redlining policies in urban centres; and 4. local laws contain strong tenant protections \citep{UGA, thomas2019state, ramiller2022displacement, redlining}. Our goal is to quantify the proportion of households moving into each Public Use Microdata Area (PUMA) within the Central Puget Sound Region that were displaced from their previous place of residence to provide policymakers and planners with a detailed picture of the state of residential displacement since 2016. Here we focus on PUMAs to ensure sample size representation to a reasonable level and to align with ACS geographies. 

In 2019 the Puget Sound Regional Council augmented the module for recent movers in their biennial Household Travel Survey (HTS). A new question on why households moved was added to complement existing data collected from households that relocated within the previous five years, such as the origin and destination of the move. However, as the HTS is primarily focused on studying transportation habits, the subsample of respondents that recently moved can be small at the PUMA-level, which introduces the risk that it may be unrepresentative at that level of disaggregation. After organizing the 2019, 2021, and 2023 HTS into two-year move cohorts (to align with the biennial nature of the survey), move cohorts within each PUMA often only have a small number of samples (see Figure \ref{fig:psrc}), which precludes many standard survey statistics techniques. 

To overcome sample size limitations in the HTS, we augment the HTS data with the ACS and AHS surveys. Specifically, we use a multilevel regression with poststratification (MRP) approach to estimate residential displacement in every PUMA in the Central Puget Sound Region for four successive two-year move cohorts. MRP partitions a target population into a granular set of subpopulations, commonly referred to as \emph{cells}, defined by a collection of categorical variables, estimates the target indicator for each subpopulation using potentially non-representative data, and then aggregates these estimates back to the target population using external population weights derived from a high-quality source \citep{gelman1997poststratification, park2004bayesian}. MRP has a rich history estimating various social and health indicators from partially or fully deficient survey data, most famously in the context of election polling \citep{park2004bayesian, wang2015forecasting}. 

Our proposed method is organized into three stages.
1. To estimate residential displacement, we first fit a Bayesian mixed-effect logistic regression model to the HTS data that includes important fixed effects like income and housing tenure as well as spatial, temporal, and survey wave random effects.
2. For each PUMA and move cohort combination, we extract the estimated risk of residential displacement from the fitted model for every subpopulation defined by the model covariates, then poststratify using population weights derived from the American Community Survey (ACS), resulting in PUMA-level residential displacement estimates for each move cohort.
3. Following the MRP procedure, to ensure consistency between the HTS- and ACS-derived PUMA-level residential displacement estimates and the AHS data, we compute the direct weighted estimates of region-wide residential displacement for each move cohort from the AHS, and apply the approach of \cite{okonek2024computationally} to ensure that aggregating our PUMA-level estimates up to the entire Central Puget Sound Region agrees with the AHS, a process known as benchmarking.

Our analysis produces sets of PUMA-level estimates of residential displacement for four consecutive move cohorts along with corresponding assessments of uncertainty. We analyse these results, identifying new insights into residential displacement in the Central Puget Sound Region in the period of 2016-2023 and opportunities for further inquiry. Our estimates are publicly available as a resource for local officials and researchers to study the patterns in displacement over space and time, and for other analyses. These results, along with code to reproduce all analyses, can be accessed at \url{https://github.com/AmeerD/displacement-mapping}. 

The remainder of this paper is organized as follows. In Section \ref{sec:data}, we describe each of the HTS, ACS, and AHS in detail and contrast their strengths and weaknesses in our context. In Section \ref{sec:methods}, we describe our MRP and benchmarking workflow used to estimate PUMA-level residential displacement. We discuss the results of our analysis in Section \ref{sec:results} and conclude with a discussion in Section \ref{sec:discussion}.
\section{Data}
\label{sec:data}

Our estimates of PUMA-level residential displacement are computed from three complementary household surveys: the Puget Sound Regional Council Household Travel Survey (HTS), the American Community Survey (ACS), and the American Housing Survey (AHS). Each survey offers a piece of the residential displacement puzzle: the HTS provides household-level residential displacement data, detailed geographic identifiers, and demographic and household characteristic variables; the ACS provides granular information on population and household characteristics; and the AHS provides a high-quality sample of residential displacement data for the Central Puget Sound Region as a whole. In this section, we discuss each survey in detail, focusing on the questions of interest, data pre-processing, and survey designs.

\subsection{Puget Sound Regional Council Household Travel Survey}

The HTS is a biennial survey conducted by the Puget Sound Regional Council that aims to collect data on the regional travel habits of residents of the Central Puget Sound Region. Respondent households are asked to provide detailed demographic, household, and geographic information (see Table \ref{tab:cov} for a summary of relevant variables), and then to record daily regional travel for up to seven consecutive days. Starting with the 2019 survey, the HTS asked \emph{why} a household left its previous home in its module on moving, in addition to when it moved to its current location. Taken together, these data offer insight into residential displacement over time by socioeconomic and demographic characteristics. Moreover, the HTS includes PUMA identifiers, allowing sub-county analyses.

\begin{table}
\centering
\caption{Summary of categorical household variables available in the HTS and ACS used for modelling.}
\label{tab:cov}
\begin{tabular}{lp{0.45\textwidth}r}
\toprule
Covariate & Categories & Number of Categories\\
\midrule
Household size & 1 person, 2 people, 3+ people & 3 \\
Household income & \textless 25,000, 25,000-49,999, 50,000-74,999, 75,000-99,999, 100,000+ & 5 \\
Number of vehicles & 0, 1, 2+ & 3 \\
Children & Yes, No & 2 \\
Ownership & Own, Rent/Other & 2 \\
Race & African American, Asian, Hispanic, White Only, Other & 5 \\
\bottomrule
\end{tabular}
\end{table}

The HTS follows a stratified design: the population of the Central Puget Sound Region is first divided based on county/city of residence, urbanicity, and whether a household falls into a certain special subpopulation of interest targeted for oversampling (e.g., in 2019 there was particular interest in City of Seattle Urban Villages) \citep{psrc2019, psrc2021, psrc2023}. Within each strata, a random sample of households was selected for participation in the study. As is common in household travel surveys, only a small fraction of invited households agreed to participate (4.52\% in the 2019 HTS for example). To account for the survey design and for the non-response, the HTS is equipped with a set of household weights. These weights are constructed starting with a design weight that accounts for systematic oversampling in certain strata, followed by a raking stage that calibrates the HTS sample to several individual and household level variables derived from the ACS, including the first four in Table \ref{tab:cov} \citep{psrc2019, psrc2021, psrc2023}.

Our goal is to estimate residential displacement at the PUMA-level for successive two-year move cohorts starting with 2016-2017 up to 2022-2023. The 2019, 2021, and 2023 HTS collectively contain the requisite granular information on \emph{when} households moved, \emph{where} they moved to, and \emph{why} they moved. 

We use the question on duration lived in the present address to determine \textit{when} a household moved and to classify those households into move cohorts. Since this question offered respondents a limited list of duration categories, this process is inexact: it provides a category for under one year and one to two years, which together capture the move cohort ending in the survey year, but then only provides categories for two to three years and three to five years, neither of which exactly match the preceding cohort. To avoid discarding a large number of respondents, we combine these two latter categories to approximate the move cohort two to four years before the survey year, although we acknowledge that this is a mismatch resulting from data limitations. Any households that have lived in the present address for longer than 5 years are not included. To determine \emph{where} households moved we georeference the respondent's home address to the relevant PUMA. Two sets of PUMA labels corresponding to the 2010 and 2020 census are available. We retain the 2010 labels for the 2016-2017, 2018-2019, and 2020-2021 move cohorts and keep the 2020 labels for the 2022-2023 move cohort. This decision is dictated by the PUMA annotations available in the ACS. Finally, regarding \emph{why} households moved, we binarize the set of reasons for moving questions into an indicator for displacement: if a household moved for income/cost related reasons, or was forced to move (by a landlord, a bank, an imminent building demolition, or some other external driver), we classify them as displaced; see Section \ref{app:question} of the supplementary materials for the specific responses classified as displaced. 

\begin{figure}[h]
\centering
\includegraphics[width=0.95\textwidth]{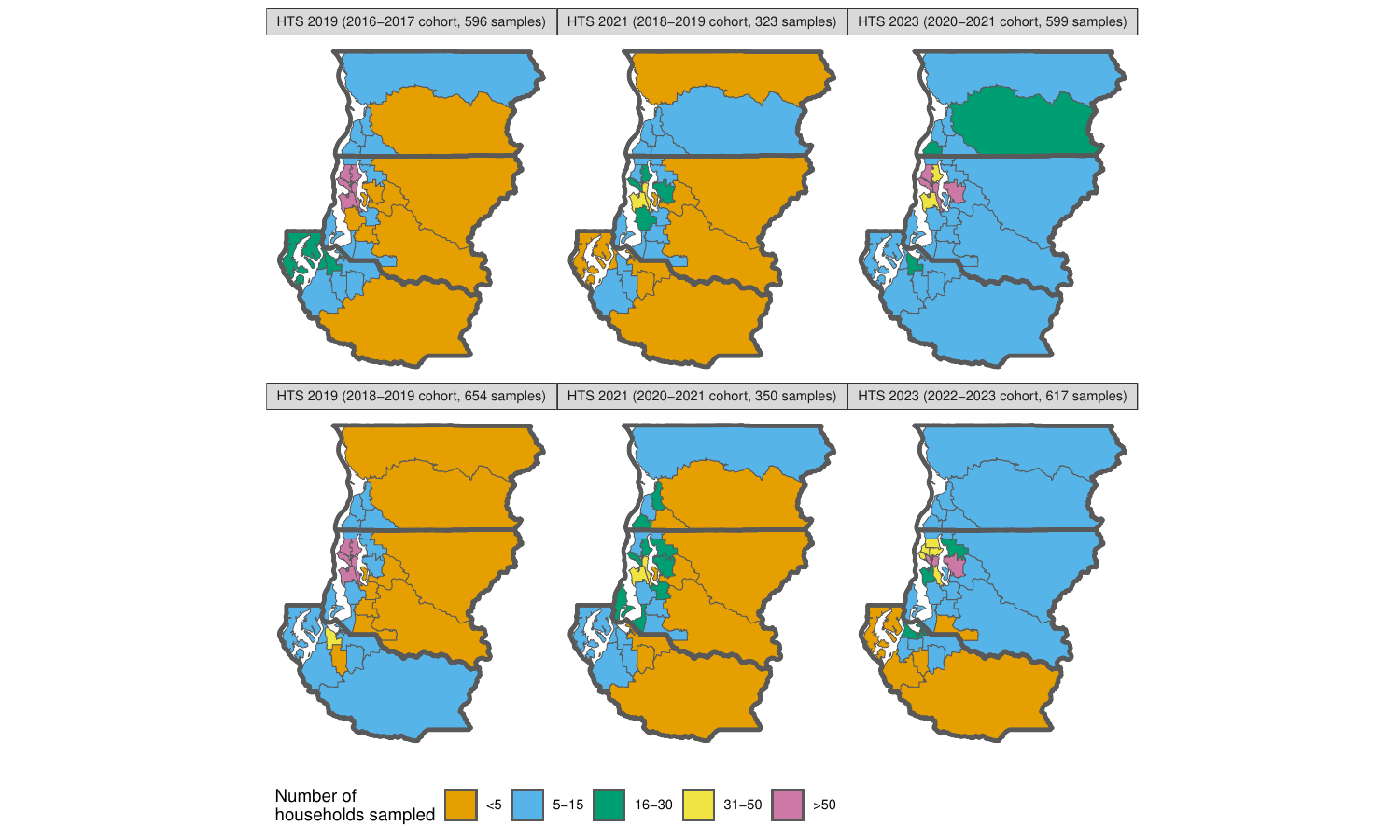}
\caption{Maps of the number of recent movers sampled by each wave of the HTS organized by move cohort and PUMA. For many move cohort and PUMA combinations, the number of samples is small, precluding direct estimation of residential displacement.}
\label{fig:psrc}
\end{figure}

Figure \ref{fig:psrc} summarises the residential displacement data available in the HTS. It displays the number of sampled households remaining after pre-processing the HTS data. Each panel contains a map of the number of households sampled within each PUMA for a given survey year and move cohort. The total number of samples in each survey year-move cohort combination are given in the panel titles. The figure clarifies why the HTS alone is not sufficient for the present task: for a large proportion of PUMAs there are only a handful of households sampled. For the 2018-2019 and 2020-2021 move cohorts, this issue is partially mitigated by the fact that we have two contributing surveys, though for many of the PUMAs farther away from population centres, the combined sample size remains small. This renders direct weighted estimation \citep{horvitz1952generalization} of residential displacement rates within each PUMA for each move cohort impractical \citep{rao2015small}. Moreover, the sample size issues in tandem with the high degree of non-response and the lack of available population information within subsets of each strata limit the applicability of standard small-area estimation models, motivating our use of MRP to integrate the HTS data with other sources. 

\subsection{American Community Survey}

The ACS is an ongoing survey conducted by the United States Census Bureau that collects demographic, economic, housing, and other information about the population of the United States \citep{ACS}. The ACS is a large survey: roughly 3.5 million households participate annually. Each year the United States Census Bureau produces summary statistics at various geographic levels for the previous year as well as for a rolling 5-year period. In addition, they release microdata containing individual-level responses that researchers can use to conduct more tailored analyses. These data products are a primary resource for studying the evolution of the American population, particularly in intercensal years. 

The ACS has a rich history of use in the MRP literature as its breadth, size, and public availability of its microdata allow users to construct high-quality population estimates for poststratification; see for example 
\cite{gao2020improving} and \cite{si2024multilevel}. In our case, we access the 2017, 2019, 2021, and 2023 1-year ACS microdata using the \texttt{tidycensus} R package \citep{tidycensus}. We first subset these records to those who moved within the previous two years. This leaves 3,434, 3,353, 3,438, and 3,465 samples for the 2016-2017, 2018-2019, 2020-2021, and 2022-2023 move cohorts, respectively. We then construct weighted estimates of the population within each PUMA corresponding to the covariates outlined in Table \ref{tab:cov}. The survey weights used in this process are provided with the microdata and were constructed by the United States Census Bureau based on the survey design followed by several adjustments for monthly variation in response patterns, non-interview, and to ensure consistency with population estimates from the United States Census Bureau's Population Estimates Program \citep{ACSweights}.

Because ACS PUMA labels are defined by the most recent census, ACS data compiled across many years may not have consistent PUMA labels. The 2017, 2019, and 2021 ACS provide PUMA labels based on the 2010 Census and the 2023 ACS provides PUMA labels based on the 2020 Census. The differences across censuses are not simply minor boundary changes; in our case, the 2010 census subdivides the Central Puget Sound Region into 29 PUMAs whereas the 2020 census subdivides the Central Puget Sound Region into 32 PUMAs. Consequently, for the 2016-2017, 2018-2019, and 2020-2021 move cohorts, we produce residential displacement rate estimates for PUMAs defined by the 2010 census and for the 2022-2023 cohort, we produce estimates for the PUMAs defined by the 2020 census. 

\subsection{American Housing Survey}

The AHS is a biennial survey funded by the United States Department of Housing and Urban Development (HUD) and implemented by the United States Census Bureau. It studies the structure of the housing stock and how it changes over time. The AHS is a longitudinal study with two branches: the national sample and the metro samples for the largest 15 metro areas in the US (a group that includes the Seattle-Tacoma-Bellevue metropolitan statistical area; i.e., the Central Puget Sound Region) \citep{AHS}. In 2015, representative samples were drawn at the national level as well as for each of the metro areas. The remainder of this subsection focuses on the Central Puget Sound Region metro sample. 

The Central Puget Sound Region sample follows a stratified design with 9 strata. Each strata corresponds to a different ``type" of housing unit; for example, renter-occupied and one unit in structure, trailer or mobile home, or vacant and two or more units in structure. Within each strata, a systematic sample was drawn from a complete list of housing units maintained by the United States Census Bureau. This list of housing units forms the initial cohort of samples. In every subsequent wave of the AHS (i.e., every two years), an additional sample drawn from the list of new builds using the same stratified design is added to the study so that the sample remains representative of the universe of households. Weights for sampled units were constructed by taking the base weight implied by the design, multiplying by a non-response adjustment, and raking to population totals sourced from other HUD and Census Population Division data products; see \cite{AHSweights} for complete details.

Our interest within the AHS is the subset of housing units occupied by recent movers; that is, occupants that moved into the unit within the past 2 years. For this subpopulation, the AHS asks respondents a series of binary questions about the reasons for moving. In keeping with the HTS, we classify recent movers that state they were forced to move by their landlord, bank or other financial institution, or the government, or that moved to reduce costs as having been displaced. For each AHS between 2017 and 2023, we compute the weighted estimate of the proportion of displaced households as well as standard errors using the provided replicate weights; the data are contained in the AHS Public Use Files \citep{AHSdata}. Table \ref{tab:ahs} summarises the size of the Central Puget Sound Region samples within the four AHS analysed and their corresponding estimates of residential displacement.

\begin{table}
\centering
\caption{Summary of American Housing Survey data.}
\label{tab:ahs}
\begin{tabular}{llrrr}
\toprule
Survey year & Move cohort & \makecell[r]{Number of recent\\movers} & \makecell[r]{Estimated Residential\\Displacement} & Standard Error\\
\midrule
2017 & 2016-2017 & 585 & 0.266 & 0.019\\
2019 & 2018-2019 & 526 & 0.212 & 0.018\\
2021 & 2020-2021 & 598 & 0.240 & 0.018\\
2023 & 2022-2023 & 451 & 0.228 & 0.021\\
\bottomrule
\end{tabular}
\end{table}

\section{Methods}
\label{sec:methods}

We estimate residential displacement for all PUMAs in the Central Puget Sound Region for each move cohort using a three-stage multilevel regression with poststratification procedure. First, we fit a spatiotemporal logistic regression model to the HTS data to estimate the probability of displacement for categories of recent movers defined by the variables in Table \ref{tab:cov} as well as PUMA and move cohort. Second, to ensure our estimates are representative of the population of recent movers, we use data on the composition of recent movers in each PUMA and move cohort combination to poststratify the estimated probabilities of displacement from our regression model. Third, we benchmark our results to ensure consistency with the aggregate AHS-derived estimates in Table \ref{tab:ahs}. We build our proposed workflow within a Bayesian framework, and produce posterior distributions of residential displacement rates in each PUMA for each move cohort. 

\subsection{A spatiotemporal model for residential displacement}

We use a Bayesian spatiotemporal logistic regression to model the log-odds of displacement as a function of a set of demographic and housing related fixed effects as well as spatial, temporal, and survey-specific random effects. The inclusion of structured random effects allows us to greatly expand the number of subpopulations available for poststratification \citep{gao2020improving}. As seen in Figure \ref{fig:psrc}, many PUMAs have limited sample sizes, precluding PUMA-specific fixed effects, a problem rectified by including a spatial random effect.

Formally, let $Y_i$ be an indicator for displacement for the $i$th recent mover in the HTS data and let $p_i$ be the corresponding probability of displacement. Then, $Y_i$ is modelled as a Bernoulli trial with probability $p_i$:
$$
Y_i \sim \text{Bernoulli}(p_i).
$$

We model $p_i$ on the logit scale as a linear combination of fixed and random effects. We consider 6 categorical or binary covariates for the fixed effects: household size, household income, number of vehicles, children, ownership, and race; see Table \ref{tab:cov} for a full breakdown of the covariates. Recognizing that the covariates may not capture all differentials in displacement across time and space, we also include temporal and spatial random effects. We include a first-order random walk model to smooth successive move cohorts. Then for space, each move cohort is assigned a discrete spatial model that captures any residual heterogeneity in the PUMAs left unexplained by the covariates. For this task, we use a BYM2 model; briefly, the BYM2 model asserts that the spatial dependence in the data can be expressed as the sum of a spatially structured effect that pools information between adjacent PUMAs as well as an unstructured effect that allows for idiosyncratic heterogeneity in each PUMA \citep{riebler2016intuitive}. Mirroring the changes in the PUMA boundaries, the spatial models corresponding to the 2016-2017, 2018-2019, and 2020-2021 move cohorts are defined with respect to the 2010 PUMA boundaries whereas the spatial model corresponding to the 2022-2023 move cohort is defined with respect to the 2020 PUMA boundaries. Finally, we include a simple independent and identically distributed random effect model for the survey effect. The survey effect is a data artifact capturing potential systematic biases in the HTS waves, and not a driver of displacement. It is thus included in the model here, but excluded downstream when constructing poststratified estimates. 

Our full model is:
\begin{align}
\text{logit}(p_i) &= \beta_0 + X_i^\top\beta + \alpha_{t[i]} + \omega_{s[i],t[i]} + \gamma_{w[i]}, \label{eq:model} \\
\alpha &\sim \text{RW1}(\sigma_\alpha^2), \nonumber \\
\omega_{\cdot,t} &\sim \text{BYM2}(\tau_{t},\phi_{t}) \quad \text{for each } t, \nonumber \\
\gamma_w &\sim N(0,\sigma_\gamma^2), \nonumber
\end{align}
where $\beta_0$ is the intercept, $X_i$ denotes the vector of indicator variables for the covariates for mover $i$, $\beta$ is the corresponding vector of regression coefficients, $t[i]$, $s[i]$, and $w[i]$ index the move cohort, PUMA, and HTS survey wave for mover $i$, respectively, $\alpha$ is the temporal random effect, $\omega_{\cdot,t}$ are the spatial effects, and $\gamma$ is the survey random effect. The random effect models are parameterised by the following hyperparameters: the temporal variance $\sigma_\alpha^2$ for $\alpha$; the marginal precision $\tau_t$ and proportion of variance attributed to space $\phi_t$ for the spatial model for each move cohort $t$; and the variance $\sigma_\gamma^2$ for the survey random effect.

\subsection{Poststratification}

The model defined in \eqref{eq:model} assigns the same probability of displacement to recent movers with the same covariate values within the same PUMA and same move cohort. We refer to each combination of covariate values as a poststratification cell. In total, there are $J=900$ cells within each PUMA and move cohort combination. In the poststratification phase, for each PUMA and move cohort combination, we aggregate the corresponding $900$ cell-level estimates into a single representative estimate of residential displacement. This is accomplished by computing their weighted average where the weights for each combination are the corresponding proportion of recent movers belonging to each cell as reported by the ACS. This ensures that the contribution of each cell to the overall estimate reflects the actual composition of recent movers. 

More precisely, we take a Monte Carlo approach. We draw $R=10,000$ samples from the posterior distributions of cell-level displacement probabilities. Let $p_{j,s,t}^{(r)}$ denote the $r$th posterior draw for cell $j$ in PUMA $s$ and move cohort $t$. Next, let $N_{j,s,t}$ denote the weighted estimate of the corresponding population derived from the ACS. Then, for each PUMA and move cohort, and for each value $r=1,\dots,R$, we compute
$$
p_{s,t}^{(r)} = \frac{\sum_{j=1}^J N_{j,s,t}p_{j,s,t}^{(r)}}{\sum_{j=1}^J N_{j,s,t}}.
$$
Here $p_{s,t}^{(r)}$ is the $r$th sample from the posterior distribution of $p_{s,t}$, the poststratified estimate of residential displacement for PUMA $s$ and move cohort $t$. These $R$ samples characterise the posterior distribution of the poststratified residential displacement probabilities. 

\subsection{Benchmarking}

In principle, the poststratification step ensures that $p_{s,t}^{(1)},\dots,p_{s,t}^{(R)}$ are representative of the rate of residential displacement within each PUMA and move cohort, and can therefore be used to compute various summaries of the posterior residential displacement distributions such as the mean, standard deviation, and quantiles. However, they are not guaranteed to be consistent with the AHS. Such a disagreement is highly undesirable: if the discrepancies are large, it can undermine the credibility of one or both sets of estimates \citep{bell2013benchmarking, zhang2020fully, okonek2024computationally}.

To protect against such a scenario, we benchmark our MRP estimates to the estimates of residential displacement for all of the Central Puget Sound Region derived from the AHS. We apply the fully Bayesian benchmarking proposals of \cite{zhang2020fully} and \cite{okonek2024computationally} that constrain the posterior using the likelihood of the benchmark; in our case, this translates to the following additional Gaussian likelihood term for the AHS estimates:
$$
p_{t}^{AHS} \sim N\left(\frac{\sum_{s}N_{s,t}p_{s,t}}{\sum_sN_{s,t}},\hat\sigma_{t}^2\right)
$$
where $p_t^{AHS}$ are the AHS-derived estimates of residential displacement in the entire Central Puget Sound Region for move cohort $t$ given in Table \ref{tab:ahs}, $N_{s,t}=\sum_{j=1}^J N_{j,s,t}$ is the estimate of the population size of move cohort $t$ in PUMA $s$, $p_{s,t}$ are the poststratified estimates of residential displacement, and $\hat\sigma_t^2$ is the replicate weight-based estimate of the variance of $p_t^{AHS}$ given in Table \ref{tab:ahs}. This can be thought of as penalizing MRP estimates that deviate too far from the benchmark.

\cite{okonek2024computationally} show that this benchmark can be incorporated with a simple post-processing step. First, they fit an unbenchmarked model and generate a large number of samples from the posterior (i.e., they produce $p_{s,t}^{(1)},\dots,p_{s,t}^{(R)}$). Then, they show that the benchmark can be implemented by ``filtering out" samples that are inconsistent with the benchmark. We use their rejection sampler to perform the filtering step. After benchmarking, we are left with a reduced set of posterior samples, which we denote by $p_{s,t}^{*(r)}$ for $r=1,\dots,R^*$ where $R^*=779$ is the number of samples that passed the filter. These reduced sets of posterior samples are the final outputs of our workflow. In the next section, we use them to compute posterior quantities of interest and ultimately to report our final estimates of residential displacement.

\subsection{Model validation}

The reliability of our modelled estimates of residential displacement is predicated on the ability of the logistic model to accurately predict the probability that a recent mover is displaced given its household characteristics. We therefore evaluate our proposed specification in \eqref{eq:model} on this task with a leave-one-area-out validation exercise. 

For each PUMA and move cohort combination, we remove all corresponding recent movers from our HTS data, and reserve these households as a validation set. We then refit our logistic model and predict the residential displacement rate for the held-out households. As comparison points, we repeat this procedure with three related model specifications. The first is a simplification that eliminates all random effects; we refer to this as the ``covariate" model. This model assumes that all six covariates adequately explain all the variability in residential displacement rates and is implemented by replacing \eqref{eq:model} with 
$$
\text{logit}(p_i) = \beta_0 + X_i^\top\beta.
$$

The second model reintroduces the temporal random effect for move cohorts as well as survey random effects, but does not include a PUMA-level spatial effect; we refer to it as the ``cohort" model. It assumes that each move cohort and each survey wave may be systematically different, but otherwise also asserts that the six covariates explain the heterogeneity in residential displacement rates. It models $\text{logit}(p_i)$ as:
$$
\text{logit}(p_i) = \beta_0 + X_i^\top\beta + \alpha_{t[i]} + \gamma_{w[i]}.
$$

The third model is the most similar to our proposed model, but rather than including a spatial model on the PUMAs, we include an independent and identically distributed random effect on PUMAs that recognizes that PUMA and move cohort combinations are nested hierarchically within each move cohort, but does not exploit their inherent spatial structure; we refer to it as the ``hierarchical" model. It is written as:
\begin{align*}
\text{logit}(p_i) &= \beta_0 + X_i^\top\beta + \alpha_{t[i]} + \omega_{s[i],t[i]} + \gamma_{w[i]},  \\
\omega_{s,t} &\sim N(0,\sigma_{t}^2) \quad \text{for all } s,t. 
\end{align*}

Performance of the four models is assessed with two metrics: 1. we compute the mean squared error (MSE) between the predicted and observed residential displacement rates, and 2. we compute coverage rates of the credible intervals at the 80\%, 90\%, and 95\% nominal levels. For the former metric, lower values are preferred, and for the latter metric, values approaching the nominal levels are preferred.

\begin{table}
\centering
\caption{Summary of leave-one-area-out cross-validation. Each row corresponds to a candidate model and the columns provide the mean squared error and coverage results.}
\label{tab:val}
\begin{tabular}{lrrrr}
\toprule
Model & MSE & Coverage (80\%) & Coverage (90\%) & Coverage (95\%)\\
\midrule
Covariate & 4.35 & 78.99 & 88.24 & 93.28\\
Cohort & 4.28 & 78.15 & 85.71 & 92.44\\
Hierarchical & 4.25 & 78.99 & 85.71 & 95.80\\
Proposed & 4.24 & 83.19 & 93.28 & 96.64\\
\bottomrule
\end{tabular}
\end{table}

Table \ref{tab:val} summarises the results of this exercise. Our model reports the lowest MSE among the four candidates. The coverage results suggest that the covariate, cohort, and hierarchical models are undercovering mildly at the 80\% and 90\% levels. By contrast, our proposed model experiences mild overcoverage at all three nominal levels, suggesting that its uncertainty intervals are better calibrated than its competitors, if a little conservative. To better understand the source of the undercoverage, in Supplement \ref{app:restables}, we disaggregate the results by move cohort and find that the three competitor models perform poorly on the 2018-2019 and 2022-2023 move cohorts, suggesting that there is meaningful unexplained heterogeneity in these cohorts that is missed by the covariates and that the simpler specifications lack the tools to capture.

\subsection{Computation}

Model parameters are estimated using the integrated nested Laplace approximation method proposed by \cite{rue2009approximate} and implemented in the \texttt{INLA} R package. All priors are set to the default uninformative priors in \texttt{INLA}. Posterior draws are generated using the \texttt{inla.posterior.sample} function. The benchmarking step is performed using the \texttt{benchmark\_sampler} function implemented in the \texttt{stbench} package \citep{okonek2024computationally}.

\section{Results}
\label{sec:results}

\subsection{PUMA-level estimates of residential displacement}

\begin{figure}
\centering
\includegraphics[width=0.95\textwidth]{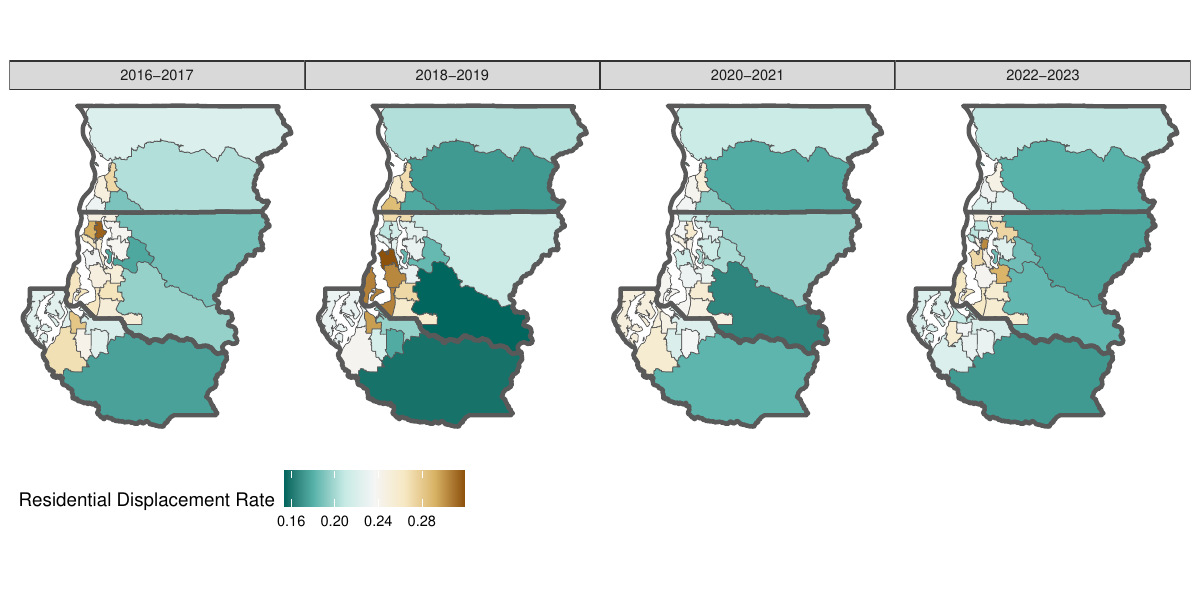}
\caption{Maps of the posterior median estimates of the residential displacement rates for each move cohort. County borders are displayed by the thick grey lines and PUMA borders are displayed by the thin grey lines. The PUMA boundaries for the 2016-2017, 2018-2019, and 2020-2021 move cohorts correspond to the 2010 census definitions whereas the PUMA boundaries for the 2022-2023 move cohort correspond to the 2020 census definitions.}
\label{fig:maps}
\end{figure}

We begin by discussing the PUMA-level estimates of residential displacement for each move cohort. Figure \ref{fig:maps} displays a series of maps of the benchmarked posterior median estimates of residential displacement (i.e., the medians of $p^{*(r)}_{s,t}$). Each map corresponds to the specific move cohort in the panel title. We note again that the PUMA boundaries for the 2022-2023 move cohort are different from the previous three cohorts. 

Across all four maps, there is a clear east-west gradient: residential displacement is estimated as higher in the urban west and lower in the suburban/rural east. This pattern persists across all three counties in the Central Puget Sound Region. Within King County specifically, the southwestern PUMAs starting in the southern reaches of Seattle and extending to the southern border of the county have elevated displacement rates. This is consistent with the understanding that communities are being displaced from the urban core of Seattle into neighbourhoods further south \citep{hwang2023moved, thomas2019state}.

The 2020-2021 move cohort is interesting as it corresponds to those who moved during the height of the COVID-19 pandemic and its associated social distancing and mitigation policies. Our estimates suggests that residential displacement flattened in this period. More research is needed to understand whether this is a consequence of pandemic era policies such as the eviction moratorium and pandemic income supports \citep{moratorium, covidbook}, a relative increase in more affluent households exploiting the weaker housing market in this period to make favourable moves (a similar finding was found by \citealt{hwang2022residential} in the Bay Area), or if this was perhaps a behavioural phenomenon driven by households resisting moves to minimize change in an already turbulent time. 

\begin{figure}
\centering
\includegraphics[width=0.95\textwidth]{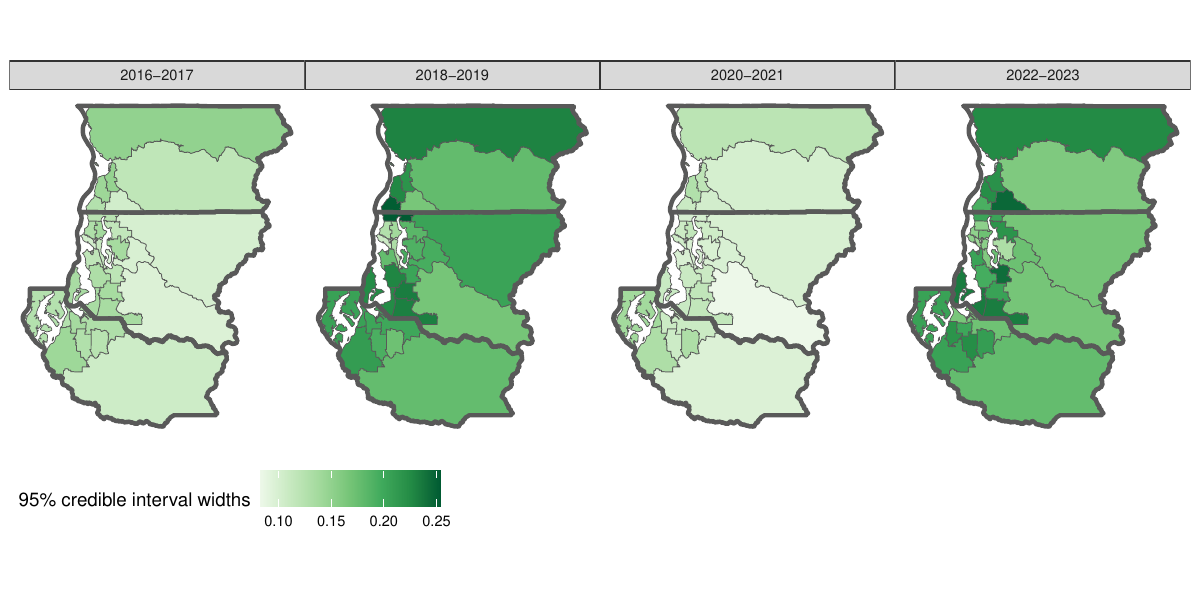}
\caption{Maps of the 95\% credible interval widths for the residential displacement rates for each move cohort. Darker shades of green correspond to wider credible intervals (i.e., greater uncertainty). County borders are displayed by the thick grey lines and PUMA borders are displayed by the thin grey lines. The PUMA boundaries for the 2016-2017, 2018-2019, and 2020-2021 move cohorts correspond to the 2010 census definitions whereas the PUMA boundaries for the 2022-2023 move cohort correspond to the 2020 census definitions.}
\label{fig:sdmaps}
\end{figure}

In addition to point estimates, for each PUMA in each move cohort, we compute 95\% credible intervals to assess the uncertainty in our benchmarked estimates. In Figure \ref{fig:sdmaps} we display maps of the widths of these intervals with one map per move cohort. The widths of the intervals range from approximately 10\%-25\%, which while not unreasonably large, implies that the point estimates should be understood as uncertain quantities and not as the ground truth. There is substantial variation in the interval widths across move cohorts with the 2016-2017 and 2020-2021 move cohorts having smaller intervals (i.e., higher precision) than the 2018-2019 and 2022-2023 move cohorts. These differences are driven by differences in the variance of the cohort-specific spatial fields, which then propagates to the residential displacement probabilities. For the 2022-2023 cohort, the higher variance may simply be a consequence of the fact that only one survey with data on this cohort is available; new data from the next wave of surveys may assist in improving precision. Alternatively, it is possible that there is fundamentally more uncertainty in the behaviour of the 2022-2023 move cohort as residents' movement patterns may have been scrambled in the aftermath of the COVID-19 pandemic. The 2018-2019 move cohort is interesting in that two surveys of data are available and yet the uncertainty remains elevated. It may be the case that there were local idiosyncrasies in the moving behaviours of this cohort that were outside the scope of the covariates included, necessitating a more flexible random effect.

For completeness, Supplement \ref{app:restables} contains tables with the posterior medians, 95\% credible interval widths, and posterior standard deviations for all PUMAs and move cohort combinations.

\subsection{Covariate estimates}

In this subsection, we explore the posterior estimates of the fixed effect coefficients in the logistic model detailed in \eqref{eq:model} fit to the HTS data. Table \ref{tab:or} summarises the posterior medians as point estimates on the odds ratio scale (relative to the baseline category holding all other variables fixed) alongside 95\% credible intervals. The table is organized according to the six covariates in Table \ref{tab:cov}.

Table \ref{tab:or} provides insight into which subpopulations are associated with higher levels of displacement. Many of the results are perhaps expected: renters have substantially higher odds of displacement than owners as renters are systematically disadvantaged in the housing market relative to owners and are further subject to the risk of being displaced by their landlord \citep{siskar2021predicting, deluca2022housing}. Our model estimates that the odds of displacement are $2.17$ times higher for renters as compared to owners that share all other characteristics (95\% credible interval: $(1.72, 2.74)$). Similarly, the finding that lower income households are at higher risk of displacement than high income households is not surprising given that financial factors are included in our definition of displacement; the degree of elevated risk scales with the income gap (see Table \ref{tab:or} for specific values). Similar findings have been found in other US cities \citep{wyly2010displacing, desmond2015forced}. Interestingly, we also find that larger households are at higher risk of displacement than individuals or pairs \citep{siskar2021predicting}. In the other direction, we find that households with two or more vehicles have lower odds of displacement relative to the corresponding baseline categories. Finally, we note that conditioning on all other variables, the role of race appears to be limited, with only the ``Other" category having a 95\% credible interval that does not cover 1; this result is consistent with previous studies including \cite{lee2020forced} and  \cite{siskar2021predicting}.

\begin{table}
\centering
\caption{Posterior medians for the odds ratios of each fixed effect relative to the baseline category holding all other variables fixed for each set of fixed effects along with the associated 95\% credible intervals. Variables related to household composition and wealth have strong associations with residential displacement rates.}
\label{tab:or}
\begin{tabular}{llrr}
\toprule
Variable & Category & Posterior Median & 95\% Credible Interval\\
\midrule
 & 2 people & 0.97 & (0.77, 1.23)\\

\multirow{-2}{*}{\raggedright\arraybackslash Household size (baseline: 1 person)} & 3+ people & 1.77 & (1.12, 2.80)\\
\cmidrule{1-4}
& \textless 25,000 & 1.70 & (1.21, 2.38)\\
 & 25,000-49,999 & 1.68 & (1.27, 2.22)\\

 & 50,000-74,999 & 1.44 & (1.09, 1.90)\\

\multirow{-4}{*}{\raggedright\arraybackslash Household income (baseline: 100,000+)} & 75,000-99,999 & 1.47 & (1.11, 1.96)\\
\cmidrule{1-4}
 & 0 & 1.09 & (0.85, 1.40)\\

\multirow{-2}{*}{\raggedright\arraybackslash Number of vehicles (baseline: 1)} & 2+ & 0.75 & (0.59, 0.96)\\
\cmidrule{1-4}
Children (baseline: No children) & Children & 0.67 & (0.44, 1.03)\\
\cmidrule{1-4}
Ownership (baseline: Own) & Rent/other & 2.17 & (1.72, 2.74)\\
\cmidrule{1-4}
 & African American & 0.88 & (0.57, 1.38)\\

 & Asian & 0.83 & (0.64, 1.08)\\

 & Hispanic & 0.77 & (0.48, 1.24)\\

\multirow{-4}{*}{\raggedright\arraybackslash Race (baseline: White Only)} & Other & 1.51 & (1.20, 1.91)\\
\bottomrule
\end{tabular}
\end{table}

\subsection{The effects of benchmarking}

\begin{figure}

\centering
\begin{subfigure}{0.35\textwidth}
    \centering
    \includegraphics[width=\textwidth]{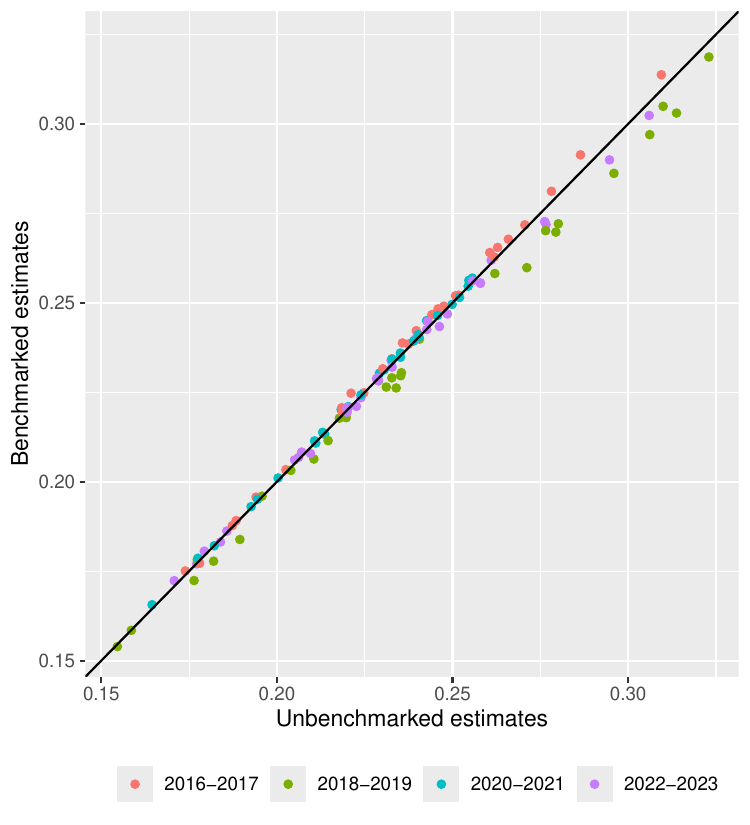}
    \caption{Posterior medians\label{fig:benchmean}}
\end{subfigure}%
~
\begin{subfigure}{0.35\textwidth}
    \centering
    \includegraphics[width=\textwidth]{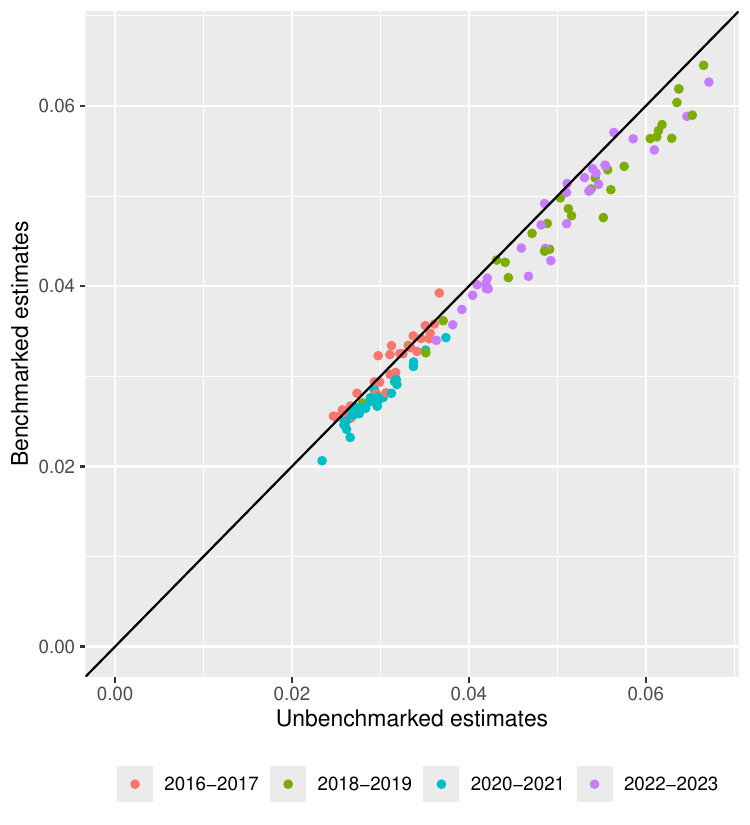}
    \caption{Posterior standard deviations\label{fig:benchsd}}
\end{subfigure}%
\caption{Results of the benchmarking stage of our residential displacement rate estimation workflow. Panel (a) plots the PUMA-level unbenchmarked posterior median estimates against the benchmarked posterior estimates, coloured by move cohort, as well as the identity line in black. The points track the line closely, implying that the unbenchmarked estimates are largely aligned with the AHS-derived estimates of residential displacement for all of the Central Puget Sound Region. Panel (b) plots the PUMA-level unbenchmarked posterior standard deviations against the benchmarked posterior standard deviations, coloured by move cohort, as well as the identity line in black. For three of the four move cohorts, the points lie below the line, implying that overall, benchmarking increases the precision of our residential displacement risk estimates.}
\end{figure}

Next, we investigate the role of benchmarking by comparing the posterior median derived from the unbenchmarked posterior samples with the posterior median derived from the benchmarked posterior samples, for each PUMA $s$ and move cohort $t$.  In Figure \ref{fig:benchmean}, each point corresponds to a PUMA and move cohort combination. The colours indicate the move cohort and the diagonal black line indicates perfect alignment. 

The fact that the unbenchmarked and benchmarked estimates are largely aligned is reassuring. It suggests that the MRP stages of our workflow are successful in correcting for any systematic biases inherent in the HTS data, an outcome that is not always possible with MRP. There does appear to be a small degree of disagreement in PUMAs with high estimated residential displacement in the 2018-2019 and 2022-2023 move cohorts. In these instances, the benchmarked estimates are slightly lower than the unbenchmarked estimates, though the difference is small, and perhaps not unexpected given the greater uncertainty in these move cohorts as shown in Figure \ref{fig:sdmaps}.

We stress that one should not interpret Figure \ref{fig:benchmean} as a suggestion that benchmarking is not important. In Figure \ref{fig:benchsd}, we compare the posterior standard deviations computed using the unbenchmarked and benchmarked posterior samples: the standard deviations of the benchmarked posteriors are substantially smaller for the 2018-2019 and 2022-2023 move cohorts, which experience the highest uncertainty. By contrast, the benchmarked posterior standard deviations are largely unchanged for the 2016-2017 cohort and slightly lower than the unbenchmarked posterior standard deviations for the 2020-2021 move cohort.

\section{Discussion}
\label{sec:discussion}

In this work, we have conducted a study of residential displacement in the Central Puget Sound Region, culminating in a complete set of PUMA-level estimates of displacement for four consecutive move cohorts alongside associated measurements of uncertainty. These high-resolution estimates offer local policymakers a tool for studying how their communities are changing, and where interventions (such as those reviewed in \citealt{chapple2023role}) to support vulnerable populations may be most needed. Unlike many displacement indicators, our estimates are not subset to gentrifying neighbourhoods nor are they an amalgamation of plausible risk factors; rather, they are based on direct measurements of displacement collected from households themselves and produced using well-understood methods from the survey statistics literature. 

A primary contribution of our work is the novel integration of the three complementary data sources through our spatiotemporal MRP workflow. The HTS, ACS, and AHS can be viewed as a three-legged stool: each survey uniquely contributes to the reliability and stability of our estimates, and something important is lost if a survey is omitted. The HTS provides highly localized information on residential displacement; the ACS ensures representativeness; and the AHS enforces consistency across sources. Moreover, all data used in this study are publicly available, thereby allowing for transparency of the estimation process. To our knowledge, this style of data integration has not previously been used in the context of residential displacement, either with the HTS in the Central Puget Sound Region, or with another local survey in another jurisdiction. 

Our methodology is transportable to other regions with similar data circumstances. The other AHS top metro areas are promising candidates for such an extension. All that is needed is a local survey with displacement information to substitute in for the HTS. In some cases, such data already exists; for example, the Detroit Metro Area Communities Study has questions on forced moves \citep{DMACS}. In cites where local displacement data are not currently available, we believe that the strategy of embedding a reasons for moving question or recent movers module into an existing survey, as was done with the HTS, could be a reasonable path forward. Our work shows that such an exercise does bear fruit. Returning to the Central Puget Sound Region context, it would also be interesting to explore producing estimates at finer geographic scales, particularly census tracts as the HTS does contain census tract labels. However, care should be taken since at such small scales, the level of uncertainty will undoubtedly increase, which may limit the utility of such estimates. 

There are some important limitations that warrant discussion. We subdivide into methodological limitations and data limitations. Starting with the former, due to sample size constraints, we are limited in the set of methods that can reasonably be applied. Standard small area estimation models, both area-level and unit-level, are not feasible. In our case, the MRP strategy was suitable, though it does not directly incorporate survey design information from the HTS. Fortunately, we are able to mitigate this issue by including several variables used in the HTS weight construction in our regression model as proxies for the design variables. A second limitation is that we do not consider uncertainty stemming from the ACS population estimates. This is fairly standard in the MRP literature, though may lead to minor understatements of the uncertainty in our estimates. 

On the data side, we again note that the small sample sizes available at the PUMA-level are the main issue. It is our hope that the present work provides a convincing argument that surveys such as the HTS are powerful tools for gathering necessary information for effective local policy decision making, and that future waves can be expanded accordingly. Beyond sample size, we note that the notion of displacement is difficult to measure precisely. Here we have classified respondents who move for income/cost reasons or those who are forced to move by a bank, landlord, or some other figure as displaced. But what about the household that moves because their community changes around them? Perhaps they can comfortably afford to stay in their current home, but choose or feel pressured to move due to negative social pressures \citep{davidson2009displacement, zukin2016gentrification}. Are they displaced? It may be worth refining displacement questions in order to capture such complex drivers of displacement in future surveys. In the present work, our definition of displacement is defined with respect to where households arrive after moving. In the context of preventing displacement, it is perhaps more interesting to study where a household is \emph{displaced from} rather than \emph{displaced to}. This is unfortunately not possible with the current data sources as the availability of previous place of residence is far sparser than current place of residence, and those who were displaced from the region altogether are omitted \citep{atkinson2000measuring}, though better record keeping in the future may unlock this line of inquiry.

To conclude, we comment that our primary goal was one of measurement: we aimed to provide reliable estimates of residential displacement at sub-county levels within the Central Puget Sound Region. From a practitioner's perspective, measurement is the first step. Ultimately, our collective goal is to mitigate residential displacement (and other harms) through policy interventions. Our work provides a means of passively monitoring changes in residential displacement over time, though disentangling the effect of specific interventions and optimizing accordingly remains a thorny issue that demands dedicated further study. 

\section*{Acknowledgments}

We would like to thank Adam Szpiro and Katherine Wilson for helpful comments on an earlier version of this work.

\bibliographystyle{natbib}
\bibliography{refs}

\newpage

\appendix
\section{Additional Results}
\label{app:restables}

\subsection{Tables of residential displacement rate estimates}

Tables \ref{tab:restable1617}--\ref{tab:restable2223} provide the complete results from our residential displacement estimation workflow. Each table contains the results for one move cohort, organized by the appropriate set of PUMAs.

\begin{table}
\centering
\caption{Full list of estimated residential displacement rates for the 2016-2017 move cohort by PUMA.}
\label{tab:restable1617}

\begin{tabular}{llrrr}
\toprule
PUMA & Move Cohort & Posterior Median & 95\% Credible Interval Width & Posterior Standard Deviation\\
\midrule
11501 & 2016-2017 & 0.281 & 0.136 & 0.034\\
11502 & 2016-2017 & 0.225 & 0.125 & 0.030\\
11503 & 2016-2017 & 0.268 & 0.141 & 0.036\\
11504 & 2016-2017 & 0.242 & 0.127 & 0.033\\
11505 & 2016-2017 & 0.219 & 0.129 & 0.032\\
\addlinespace
11506 & 2016-2017 & 0.225 & 0.123 & 0.029\\
11507 & 2016-2017 & 0.175 & 0.106 & 0.025\\
11601 & 2016-2017 & 0.291 & 0.129 & 0.034\\
11602 & 2016-2017 & 0.314 & 0.138 & 0.036\\
11603 & 2016-2017 & 0.263 & 0.104 & 0.026\\
\addlinespace
11604 & 2016-2017 & 0.252 & 0.111 & 0.027\\
11605 & 2016-2017 & 0.234 & 0.118 & 0.028\\
11606 & 2016-2017 & 0.247 & 0.123 & 0.030\\
11607 & 2016-2017 & 0.239 & 0.115 & 0.028\\
11608 & 2016-2017 & 0.238 & 0.136 & 0.034\\
\addlinespace
11609 & 2016-2017 & 0.177 & 0.099 & 0.025\\
11610 & 2016-2017 & 0.252 & 0.119 & 0.028\\
11611 & 2016-2017 & 0.248 & 0.130 & 0.032\\
11612 & 2016-2017 & 0.264 & 0.134 & 0.033\\
11613 & 2016-2017 & 0.265 & 0.135 & 0.033\\
\addlinespace
11614 & 2016-2017 & 0.252 & 0.135 & 0.033\\
11615 & 2016-2017 & 0.196 & 0.096 & 0.026\\
11616 & 2016-2017 & 0.188 & 0.100 & 0.026\\
11701 & 2016-2017 & 0.232 & 0.126 & 0.033\\
11702 & 2016-2017 & 0.249 & 0.142 & 0.033\\
\addlinespace
11703 & 2016-2017 & 0.272 & 0.148 & 0.035\\
11704 & 2016-2017 & 0.189 & 0.104 & 0.026\\
11705 & 2016-2017 & 0.203 & 0.117 & 0.029\\
11706 & 2016-2017 & 0.221 & 0.149 & 0.039\\
\bottomrule
\end{tabular}

\end{table}

\begin{table}
\centering
\caption{Full list of estimated residential displacement rates for the 2018-2019 move cohort by PUMA.}
\label{tab:restable1819}

\begin{tabular}{llrrr}
\toprule
PUMA & Move Cohort & Posterior Median & 95\% Credible Interval Width & Posterior Standard Deviation\\
\midrule
11501 & 2018-2019 & 0.297 & 0.201 & 0.052\\
11502 & 2018-2019 & 0.230 & 0.205 & 0.053\\
11503 & 2018-2019 & 0.240 & 0.209 & 0.053\\
11504 & 2018-2019 & 0.218 & 0.191 & 0.048\\
11505 & 2018-2019 & 0.196 & 0.200 & 0.050\\
\addlinespace
11506 & 2018-2019 & 0.178 & 0.171 & 0.046\\
11507 & 2018-2019 & 0.158 & 0.177 & 0.043\\
11601 & 2018-2019 & 0.206 & 0.127 & 0.033\\
11602 & 2018-2019 & 0.218 & 0.139 & 0.036\\
11603 & 2018-2019 & 0.228 & 0.108 & 0.027\\
\addlinespace
11604 & 2018-2019 & 0.226 & 0.130 & 0.033\\
11605 & 2018-2019 & 0.319 & 0.177 & 0.047\\
11606 & 2018-2019 & 0.270 & 0.253 & 0.062\\
11607 & 2018-2019 & 0.226 & 0.183 & 0.049\\
11608 & 2018-2019 & 0.230 & 0.189 & 0.048\\
\addlinespace
11609 & 2018-2019 & 0.184 & 0.194 & 0.044\\
11610 & 2018-2019 & 0.229 & 0.202 & 0.051\\
11611 & 2018-2019 & 0.303 & 0.227 & 0.060\\
11612 & 2018-2019 & 0.305 & 0.224 & 0.056\\
11613 & 2018-2019 & 0.272 & 0.233 & 0.059\\
\addlinespace
11614 & 2018-2019 & 0.258 & 0.231 & 0.056\\
11615 & 2018-2019 & 0.154 & 0.166 & 0.043\\
11616 & 2018-2019 & 0.211 & 0.204 & 0.051\\
11701 & 2018-2019 & 0.286 & 0.251 & 0.065\\
11702 & 2018-2019 & 0.260 & 0.226 & 0.057\\
\addlinespace
11703 & 2018-2019 & 0.270 & 0.217 & 0.057\\
11704 & 2018-2019 & 0.178 & 0.165 & 0.041\\
11705 & 2018-2019 & 0.172 & 0.178 & 0.044\\
11706 & 2018-2019 & 0.203 & 0.229 & 0.058\\
\bottomrule
\end{tabular}

\end{table}

\begin{table}
\centering
\caption{Full list of estimated residential displacement rates for the 2020-2021 move cohort by PUMA.}
\label{tab:restable2021}

\begin{tabular}{llrrr}
\toprule
PUMA & Move Cohort & Posterior Median & 95\% Credible Interval Width & Posterior Standard Deviation\\
\midrule
11501 & 2020-2021 & 0.245 & 0.108 & 0.028\\
11502 & 2020-2021 & 0.250 & 0.138 & 0.034\\
11503 & 2020-2021 & 0.256 & 0.130 & 0.031\\
11504 & 2020-2021 & 0.220 & 0.107 & 0.027\\
11505 & 2020-2021 & 0.221 & 0.108 & 0.026\\
\addlinespace
11506 & 2020-2021 & 0.236 & 0.130 & 0.033\\
11507 & 2020-2021 & 0.182 & 0.096 & 0.023\\
11601 & 2020-2021 & 0.235 & 0.108 & 0.026\\
11602 & 2020-2021 & 0.257 & 0.116 & 0.029\\
11603 & 2020-2021 & 0.247 & 0.105 & 0.026\\
\addlinespace
11604 & 2020-2021 & 0.241 & 0.099 & 0.026\\
11605 & 2020-2021 & 0.214 & 0.098 & 0.025\\
11606 & 2020-2021 & 0.211 & 0.100 & 0.024\\
11607 & 2020-2021 & 0.224 & 0.106 & 0.027\\
11608 & 2020-2021 & 0.213 & 0.099 & 0.027\\
\addlinespace
11609 & 2020-2021 & 0.201 & 0.098 & 0.025\\
11610 & 2020-2021 & 0.230 & 0.109 & 0.027\\
11611 & 2020-2021 & 0.231 & 0.109 & 0.028\\
11612 & 2020-2021 & 0.239 & 0.109 & 0.027\\
11613 & 2020-2021 & 0.255 & 0.116 & 0.030\\
\addlinespace
11614 & 2020-2021 & 0.246 & 0.115 & 0.028\\
11615 & 2020-2021 & 0.166 & 0.083 & 0.021\\
11616 & 2020-2021 & 0.195 & 0.098 & 0.025\\
11701 & 2020-2021 & 0.234 & 0.109 & 0.028\\
11702 & 2020-2021 & 0.241 & 0.127 & 0.029\\
\addlinespace
11703 & 2020-2021 & 0.251 & 0.118 & 0.029\\
11704 & 2020-2021 & 0.193 & 0.103 & 0.025\\
11705 & 2020-2021 & 0.179 & 0.101 & 0.025\\
11706 & 2020-2021 & 0.211 & 0.120 & 0.032\\
\bottomrule
\end{tabular}

\end{table}

\begin{table}
\centering
\caption{Full list of estimated residential displacement rates for the 2022-2023 move cohort by PUMA.}
\label{tab:restable2223}

\begin{tabular}{llrrr}
\toprule
PUMA & Move Cohort & Posterior Median & 95\% Credible Interval Width & Posterior Standard Deviation\\
\midrule
23301 & 2022-2023 & 0.247 & 0.202 & 0.051\\
23302 & 2022-2023 & 0.177 & 0.167 & 0.039\\
23303 & 2022-2023 & 0.273 & 0.217 & 0.055\\
23304 & 2022-2023 & 0.223 & 0.131 & 0.034\\
23305 & 2022-2023 & 0.186 & 0.173 & 0.041\\
\addlinespace
23306 & 2022-2023 & 0.183 & 0.167 & 0.041\\
23307 & 2022-2023 & 0.255 & 0.232 & 0.062\\
23308 & 2022-2023 & 0.262 & 0.233 & 0.057\\
23309 & 2022-2023 & 0.256 & 0.204 & 0.053\\
23310 & 2022-2023 & 0.290 & 0.242 & 0.059\\
\addlinespace
23311 & 2022-2023 & 0.240 & 0.194 & 0.047\\
23312 & 2022-2023 & 0.272 & 0.203 & 0.053\\
23313 & 2022-2023 & 0.243 & 0.160 & 0.040\\
23314 & 2022-2023 & 0.302 & 0.154 & 0.040\\
23315 & 2022-2023 & 0.243 & 0.142 & 0.036\\
\addlinespace
23316 & 2022-2023 & 0.206 & 0.151 & 0.037\\
23317 & 2022-2023 & 0.228 & 0.171 & 0.047\\
23318 & 2022-2023 & 0.207 & 0.152 & 0.040\\
25301 & 2022-2023 & 0.208 & 0.165 & 0.044\\
25302 & 2022-2023 & 0.234 & 0.200 & 0.050\\
\addlinespace
25303 & 2022-2023 & 0.219 & 0.171 & 0.043\\
25304 & 2022-2023 & 0.172 & 0.177 & 0.044\\
25305 & 2022-2023 & 0.229 & 0.208 & 0.051\\
25306 & 2022-2023 & 0.229 & 0.222 & 0.052\\
25307 & 2022-2023 & 0.256 & 0.217 & 0.053\\
\addlinespace
25308 & 2022-2023 & 0.221 & 0.205 & 0.051\\
26101 & 2022-2023 & 0.208 & 0.224 & 0.056\\
26102 & 2022-2023 & 0.181 & 0.161 & 0.040\\
26103 & 2022-2023 & 0.221 & 0.245 & 0.063\\
26104 & 2022-2023 & 0.221 & 0.192 & 0.049\\
\addlinespace
26105 & 2022-2023 & 0.232 & 0.222 & 0.053\\
26106 & 2022-2023 & 0.245 & 0.219 & 0.051\\
\bottomrule
\end{tabular}

\end{table}

\subsection{Additional model validation results}

Table \ref{tab:cohortval} provides coverage results from the leave-one-area-out cross-validation exercise, disaggregated by move cohort.

\begin{table}
\centering
\caption{Coverage results from the leave-one-area-out cross-validation exercise disaggregated by move cohort. Each row corresponds to a candidate model and move cohort and the columns provide the and coverage results.}
\label{tab:cohortval}
\begin{tabular}{llrrr}
\toprule
Model & Move Cohort & Coverage (80\%) & Coverage (90\%) & Coverage (95\%)\\
\midrule
Covariate & 2016-2017 & 89.66 & 93.10 & 96.55\\
Cohort & 2016-2017 & 89.66 & 93.10 & 96.55\\
Hierarchical & 2016-2017 & 89.66 & 93.10 & 96.55\\
Proposed & 2016-2017 & 82.76 & 93.10 & 100.00\\
\addlinespace
Covariate & 2018-2019 & 72.41 & 79.31 & 86.21\\
Cohort & 2018-2019 & 75.86 & 75.86 & 82.76\\
Hierarchical & 2018-2019 & 75.86 & 79.31 & 96.55\\
Proposed & 2018-2019 & 79.31 & 89.66 & 96.55\\
\addlinespace
Covariate & 2020-2021 & 82.76 & 93.10 & 96.55\\
Cohort & 2020-2021 & 82.76 & 89.66 & 96.55\\
Hierarchical & 2020-2021 & 82.76 & 89.66 & 96.55\\
Proposed & 2020-2021 & 86.21 & 96.55 & 96.55\\
\addlinespace
Covariate & 2022-2023 & 71.88 & 87.50 & 93.75\\
Cohort & 2022-2023 & 65.62 & 84.38 & 93.75\\
Hierarchical & 2022-2023 & 68.75 & 81.25 & 93.75\\
Proposed & 2022-2023 & 84.38 & 93.75 & 93.75\\
\bottomrule
\end{tabular}
\end{table}
\section{HTS reasons for moving question}
\label{app:question}

The HTS offers respondents the following options to answer the question: ``Which of the following were important factors in your decision to move away from your previous residence? Select all that apply." If any of the bolded options are selected, the respondent is classified as having been displaced.
\begin{itemize}
    \item \textbf{Increase in rent or housing costs, could no longer afford previous place}
    \item \textbf{Change in income or finances, could no longer afford previous place}
    \item Friends, family, or cultural community left or were leaving the area
    \item Change in who you live with (e.g., move out on your own, getting married/divorced)
    \item Needed more space
    \item Needed less space
    \item To have better access to work (e.g., better commute or take new job)
    \item To have better access to recreation, restaurants, shops, and other amenities
    \item Employer’s telework policy removed need to live in previous residence 
    \item Access to a different K-12 school
    \item Concerns about safety or crime
    \item To upgrade to a better-quality home or to stop renting and buy a home
    \item \textbf{Forced to move out (e.g., building demolished or renovated, asked to leave by landlord, foreclosure)}
    \item Other reason
    \item Prefer not to answer
\end{itemize}

\end{document}